\documentclass[12pt]{article}
\usepackage{graphicx}
\usepackage[margin=1.0in]{geometry}
\usepackage{amssymb}
\usepackage[affil-it]{authblk}
\usepackage{hyperref}

\makeatletter
\def\@fnsymbol#1{\ensuremath{\ifcase#1\or \dagger\or *\or \ddagger\or
   \mathsection\or \mathparagraph\or \|\or **\or \dagger\dagger
   \or \ddagger\ddagger \else\@ctrerr\fi}}
\makeatother
\newcommand*\samethanks[1][\value{footnote}]{\footnotemark[#1]}

\begin{document}

\title{Measuring muon tracks in Baikal-GVD\\ using a fast reconstruction algorithm}


\author[a]{V.A.~Allakhverdyan}
\author[b]{A.D.~Avrorin}
\author[b]{A.V.~Avrorin}
\author[b]{V.M.~Aynutdinov}
\author[c]{R.~Bannasch}
\author[d]{Z.~Barda\v{c}ov\'{a}}
\author[a]{I.A.~Belolaptikov}
\author[a]{I.V.~Borina}
\author[a]{V.B.~Brudanin\thanks{Deceased.	}}
\author[e]{N.M.~Budnev}
\author[a]{V.Y.~Dik}
\author[b]{G.V.~Domogatsky}
\author[b]{A.A.~Doroshenko}
\author[a,d]{R.~Dvornick\'{y}}
\author[e]{A.N.~Dyachok}
\author[b]{Zh.-A.M.~Dzhilkibaev}
\author[d]{E.~Eckerov\'{a}}
\author[a]{T.V.~Elzhov}
\author[f]{L.~Fajt}
\author[g]{S.V.~Fialkovski\samethanks}
\author[e]{A.R.~Gafarov}
\author[b]{K.V.~Golubkov}
\author[a]{N.S.~Gorshkov}
\author[e]{T.I.~Gress}
\author[a]{M.S.~Katulin}
\author[c]{K.G.~Kebkal}
\author[c]{O.G.~Kebkal}
\author[a]{E.V.~Khramov}
\author[a]{M.M.~Kolbin}
\author[a]{K.V.~Konischev}
\author[h]{K.A.~Kopa\'{n}ski}
\author[a]{A.V.~Korobchenko}
\author[b]{A.P.~Koshechkin}
\author[i]{V.A.~Kozhin}
\author[a]{M.V.~Kruglov}
\author[b]{M.K.~Kryukov}
\author[g]{V.F.~Kulepov}
\author[h]{Pa.~Malecki}
\author[a]{Y.M.~Malyshkin}
\author[b]{M.B.~Milenin}
\author[e]{R.R.~Mirgazov}
\author[a]{D.V.~Naumov}
\author[a]{V.~Nazari}
\author[h]{W.~Noga}
\author[b]{D.P.~Petukhov}
\author[a]{E.N.~Pliskovsky}
\author[j]{M.I.~Rozanov}
\author[a]{V.D.~Rushay}
\author[e]{E.V.~Ryabov}
\author[b]{G.B.~Safronov\thanks{Corresponding author.}}
\author[a]{B.A.~Shaybonov}
\author[b]{M.D.~Shelepov}
\author[a,d,f]{F.~\v{S}imkovic}
\author[a]{A.E. Sirenko}
\author[i]{A.V.~Skurikhin}
\author[a]{A.G.~Solovjev}
\author[a]{M.N.~Sorokovikov}
\author[f]{I.~\v{S}tekl}
\author[b]{A.P.~Stromakov}
\author[a]{E.O.~Sushenok}
\author[b]{O.V.~Suvorova}
\author[e]{V.A.~Tabolenko}
\author[e]{B.A.~Tarashansky}
\author[a]{Y.V.~Yablokova}
\author[c]{S.A.~Yakovlev}
\author[b]{D.N.~Zaborov\samethanks[2]}

\affil[a]{Joint Institute for Nuclear Research, Dubna, Russia}
\affil[b]{Institute for Nuclear Research, Russian Academy of Sciences, Moscow, Russia}
\affil[c]{EvoLogics GmbH, Berlin, Germany}
\affil[d]{Comenius University, Bratislava, Slovakia}
\affil[e]{Irkutsk State University, Irkutsk, Russia}
\affil[f]{Czech Technical University in Prague, Prague, Czech Republic}
\affil[g]{Nizhny Novgorod State Technical University, Nizhny Novgorod, Russia}
\affil[h]{Institute of Nuclear Physics of Polish Academy of Sciences (IFJ~PAN), Krak\'{o}w, Poland}
\affil[i]{Skobeltsyn Institute of Nuclear Physics, Moscow State University, Moscow, Russia}
\affil[j]{St.~Petersburg State Marine Technical University, St.Petersburg, Russia}

\date{}

\maketitle

\begin{abstract}
The Baikal Gigaton Volume Detector (Baikal-GVD) is a km$^3$-scale neutrino detector currently under construction in Lake Baikal, Russia.
The detector consists of several thousand optical sensors arranged on vertical strings, with 36 sensors per string. 
The strings are grouped into clusters of 8 strings each.
Each cluster can operate as a stand-alone neutrino detector.
The detector layout is optimized for the measurement of astrophysical neutrinos with energies of $\sim$ 100 TeV and above.
Events resulting from charged current interactions of muon (anti-)neutrinos will have a track-like topology in Baikal-GVD.
A fast $\chi^2$-based reconstruction algorithm has been developed to reconstruct such track-like events.
The algorithm has been applied to data collected in 2019 from the first five operational clusters of Baikal-GVD,
resulting in observations of both downgoing atmospheric muons and upgoing atmospheric neutrinos.
This serves as an important milestone towards experimental validation of the Baikal-GVD design.
The analysis is limited to single-cluster data, favoring nearly-vertical tracks.

\end{abstract}

\section{Introduction}

The Baikal Gigaton Volume Detector (Baikal-GVD) is a cubic-kilometer scale underwater neutrino detector currently under construction in Lake Baikal (Russia) at
a location with a depth of 1366~m \cite{Baikal-GVD}.
The detector uses 10-inch photomultiplier tubes (PMTs) to detect the Cherenkov light
from charged particles produced in neutrino interactions.
The detector elements are arranged along vertical strings which are in turn arranged in heptagonal clusters.
Each cluster has its own connection to the shore station and acts as an independent detector.
Eight clusters, with a total of 2304 PMTs, have already been deployed.
Six more clusters are scheduled for deployment in the next three years.
Baikal-GVD is aimed to provide observations of the TeV--PeV neutrino sky with a sensitivity similar to that of IceCube \cite{IceCube} and KM3NeT-ARCA \cite{KM3NeT_LoI}, with a complementary field of view.
Simultaneous operation of Baikal-GVD with the other neutrino telescopes allows for continuous monitoring of transient phenomena over the full sky.

A large flux of muons produced in the atmosphere above the detector
by high energy cosmic rays passes through the detector.
These downward-going muons form the dominant background for neutrino observations.
Additionally, a flux of atmospheric neutrinos, arriving from all directions, 
represents a quasi-irreducible background for astrophysical neutrino searches.
Both backgrounds need to be studied and understood 
to ensure reliable measurements of astrophysical neutrinos. 
To this end, a $\chi^2$-based track reconstruction algorithm has been developed.
The algorithm has been optimized for the atmospheric neutrino detection.
Using this algorithm, a first analysis of muon tracks recorded by Baikal-GVD has been performed.
This paper presents the algorithm, the analysis, and the obtained results.

The present analysis is limited to single-cluster data.
This is motivated by the fact that over 90\% of all triggered events in Baikal-GVD are single-cluster events.
Furthermore, the analysis of single-cluster events is independent of the inter-cluster time synchronization and calibration which themselves require dedicated cross checks and validation studies.
It should be noted, however, that the single-cluster analysis has a limited angular coverage, favoring nearly-vertical tracks.
An analysis of multi-cluster events will be presented elsewhere.

The $\chi^2$-based reconstruction algorithm is well suited for online data analysis thanks to its high speed.
It is currently integrated with the Baikal-GVD data processing system.
Other track reconstruction algorithms are currently being developed for Baikal-GVD, 
which promise improvements in neutrino detection efficiency and angular resolution relative to the algorithm described here, albeit at the cost of higher CPU requirements.

The paper is organized as follows. 
The Baikal-GVD detector is described in Section~\ref{sect:detector}.
The dataset and Monte Carlo simulations are described in Sections~\ref{sect:dataset} and \ref{sect:mc}, respectively.
Section \ref{sect:reco} presents the track reconstruction algorithm developed for this analysis.
Section~\ref{sect:analysis} describes the data analysis procedure and the obtained results.
The performance of the reconstruction algorithm is discussed in Sect.~\ref{sect:performance}.
Section~\ref{sect:conclusion} concludes the paper.

\section{The Baikal-GVD detector}
\label{sect:detector}
\begin{figure}
  \centering
  \includegraphics[height=8.5cm]{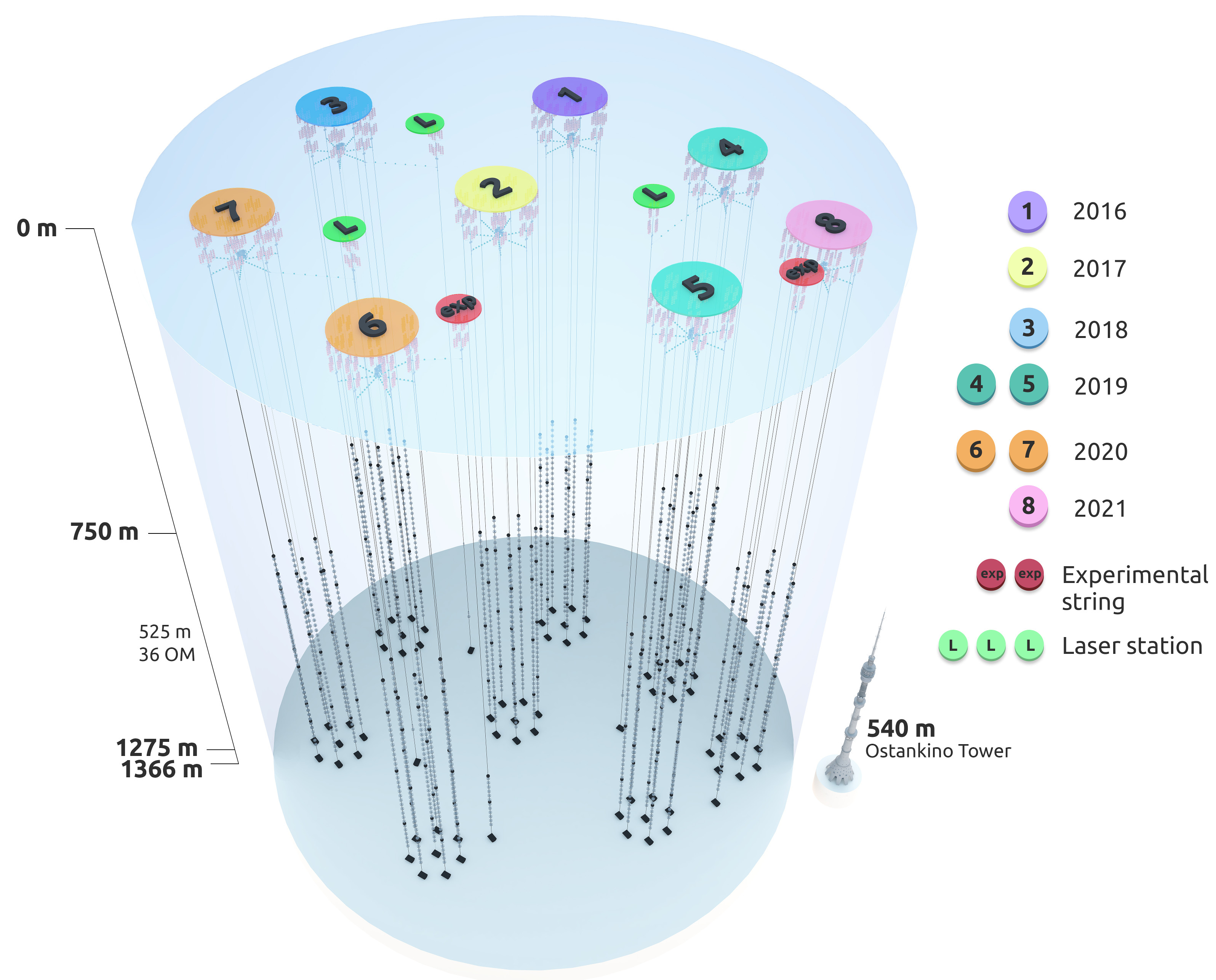}
  \includegraphics[height=8.5cm]{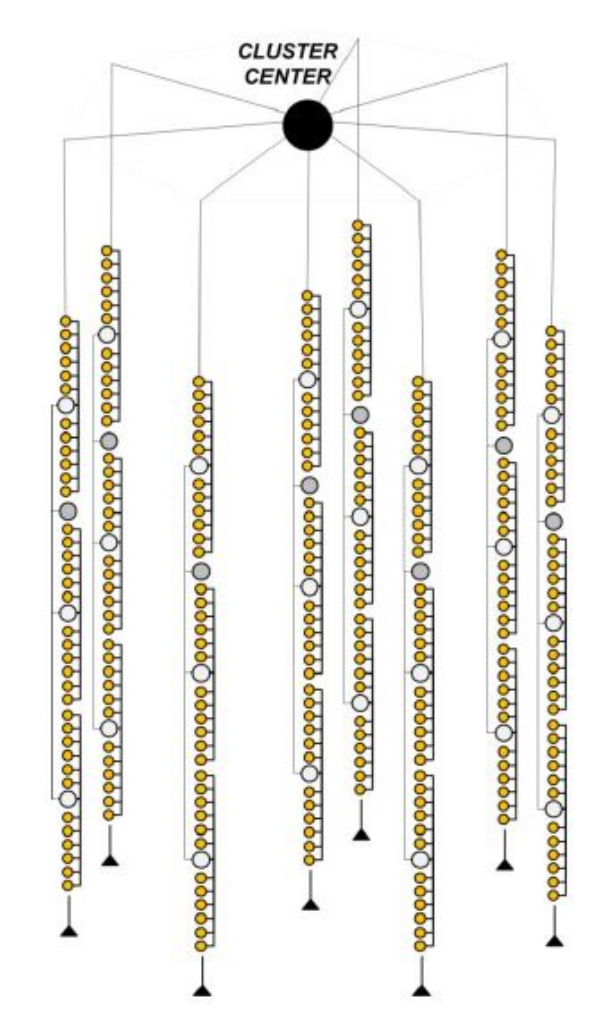}
  \caption{Left: Schematic view of the Baikal-GVD detector.
           The yearly progression of the detector deployment is shown in the legend.
           Right: The Baikal-GVD cluster layout (vertical scale compressed).}
  \label{fig:baikal_gvd}
\end{figure}

The Baikal-GVD detector is installed 3.6 km offshore in the southern basin of Lake Baikal at 51$^\circ$\,46'\,N 104$^\circ$\,24'\,E.
The lake depth at the detector location is nearly constant at 1366--1367~m below the nominal surface level of the lake.
The detector elements are arranged along vertical strings, each of which is anchored to the bottom of the lake and kept taut by a bunch of buoys at the top (see Fig.~\ref{fig:baikal_gvd}).
Each string holds 36 optical modules.
The optical module (OM) comprises a 10-inch high-quantum-efficiency PMT 
(Hamamatsu R7081-100), a high voltage unit and front-end electronics, all together enclosed in a pressure-resistant glass sphere.
The OM is also equipped with calibration LEDs and various digital sensors, including an accelerometer/tiltmeter, a compass, a pressure sensor, a humidity sensor, and two temperature sensors.
The OMs are installed with 15 m vertical spacing, for a total active string length of 525 m, starting 90~m above the lakebed.
The PMT photocathodes are oriented vertically downwards.
The OMs are mechanically attached to a load-carrying cable.
Additionally, each string has four electronics modules, also housed inside glass spheres.
Three of these modules are ``Section Modules'', each of which serves a group of 12 OMs.
The Section Modules provide power to the OMs and digitize the PMT signals with a 5~ns resolution.
The OMs are connected to their respective Section Modules via electric cables which are laid along the load-carrying cable.
The fourth electronics module is the ``String Module''.
The String Module acts as a hub for power distribution and communication with the Section Modules.
Additionally, the string holds hydrophones for acoustic monitoring of the PMT positions \cite{Baikal_positioning} and LED beacons for detector calibration \cite{Baikal_calibration,Baikal_time_calibration}.

The strings are grouped in clusters, with 8 strings per cluster, as shown in Fig.~\ref{fig:baikal_gvd}.
The positions of the strings on the lakebed form a heptagon with one central string and seven peripheral strings, with an average horizontal spacing between the strings of $\approx$ 60 m.
The eight strings are connected to the ``Cluster Center'' which is installed on the central string of the cluster at a shallow depth.
The Cluster Center comprises a system of four electronics modules.
It is connected to the shore station via a dedicated electro-optical cable.
The Cluster Center is responsible for distributing the time synchronization signals to the individual strings, cluster-level triggering and data transmission to shore.
The standard trigger condition requires two neighboring channels within the same section (of 12 OMs) to be hit within a 100 ns time window
with minimal requirements for the hit amplitudes: $a > A_{1}$ for one of the hits and $a > A_{2}$ for the other.
The amplitude thresholds $A_{1}$ and $A_{2}$ 
are set on a channel by channel basis and can be adjusted as needed to accommodate changes in the data taking conditions.
For the dataset analyzed in this work, $A_{1}$ = 3.5 photo-electrons (p.e.) and $A_{2}$ = 1.7 p.e., on average. 
Once this trigger condition is met for any of the Section Modules, a 5~$\mu$s event time frame is read out from all the Section Modules of the cluster.
Thus each cluster can operate as a stand-alone neutrino detector.
The effective volume of a GVD cluster for cascade-like neutrino events with energy above 100 TeV is estimated as 0.05 km$^3$ \cite{Baikal_cascades}.

The clusters are arranged on the lakebed in a hexagonal pattern, with a $\approx 300$~m distance between the cluster centers.
A common synchronization clock allows for subsequent merging of the physics event data collected from the different clusters.
Additional technical strings equipped with high-power pulsed lasers are installed in-between the GVD clusters. These are used for detector calibration \cite{Baikal_time_calibration} and light propagation studies \cite{Balkanov1999}.
The lake is covered with thick ice (up to $\approx$ 1~m) from February to mid-April,
providing a convenient solid platform for detector deployment and maintenance operations.

According to a study made with a specialized device, the light absorption length in the deep lake water reaches maximal values, $\approx$ 24~m, at a wavelength of 488 nm \cite{Baikal_optical_water_properties}.
The effective light scattering length is $\approx$ 480~m (at 475 nm; see \cite{Baikal_optical_water_properties} for details).
Both the absorption and scattering characteristics show variations with depth and over time.

The optical modules detect the Cherenkov light from secondary charged particles resulting from neutrino interactions.
The times of the pulses are used to reconstruct the neutrino direction, and the integrated charges (or amplitudes) provide a measure of the neutrino energy.
The detector layout is optimized for the measurement of astrophysical neutrinos in the TeV--PeV energy range.
Events resulting from charged current (CC) interactions of muon (anti-)neutrinos will have a track-like topology, while the CC interactions of the other neutrino flavors and  neutral current (NC) interactions of all flavors will typically be observed as nearly point-like events.
Hence the observed neutrino events are classified into two event classes: tracks and cascades.

The first cluster of Baikal-GVD was deployed in 2016. 
Two more clusters were added in 2017 and 2018, followed by two more in 2019, another two in 2020, and one more in 2021.
As of April 2021, the detector consists of 8 clusters, 
occupying a water volume of $\approx$ 0.4 km$^3$.
As it stands, Baikal-GVD is currently the largest neutrino telescope in the Northern Hemisphere.
The construction plan for the period from 2022 to 2024 anticipates the deployment of six additional GVD clusters.

All Baikal-GVD clusters generally show stable operation.
Occasional failures of individual optical or electronics modules, e.g. due to water leaks, are fixed during the regular winter campaigns.
Each detector string can be recovered and re-deployed without the need to recover the whole
cluster.

\section{The dataset}
\label{sect:dataset}
In this work we use a dataset collected from the first five operational clusters of Baikal-GVD in the early part of the 2019 season, between April 1 and June 30.
This period is characterized by relatively quiet optical noise levels
(see \cite{baikal_optical_noise,baikal_optical_noise_monitoring} for a review of the optical noise conditions at the Baikal-GVD site).
The average measured rate of noise hits observed by the OMs in this period ranges from $\approx 15$ kHz (at the bottommost detector storeys) to $\approx 50$ kHz (at the topmost storeys).
The total single-cluster-equivalent livetime of the dataset is $\approx 323$~days.
The livetime and number of operating optical modules for each cluster are summarized in Table~\ref{table:dataset}.
As will be shown below, this dataset is sufficient to demonstrate the observation of atmospheric neutrinos.
Additional studies are in progress to ensure the analysis stability under high optical noise conditions, as observed in July--September 2019 and parts of 2020, and to extend the analysis to the full dataset already collected by Baikal-GVD.
In this work the data from each cluster are analyzed separately.

\begin{table}[h!]
  \begin{center}
    \begin{tabular}{|c|c|c|}
      \hline
      \textbf{GVD cluster} & \textbf{Number of active OMs} & \textbf{Dataset duration, days}\\
      \hline
      1 & 270 & 68\\
      2 & 273 & 72\\
      3 & 288 & 74\\
      4 & 288 & 61\\
      5 & 288 & 47\\
      \hline
      combined & 1407 & 323\\
      \hline
    \end{tabular}
    \caption{Basic characteristics of the dataset used in this work
(April 1 -- June 30, 2019).
}
    \label{table:dataset}
  \end{center}
\end{table}

\section{Monte Carlo simulations}
\label{sect:mc}
Dedicated Monte Carlo (MC) simulations have been performed to estimate the expected rate of atmospheric muon and atmospheric neutrino events in Baikal-GVD.
Only track-like events are considered in the present study.
The track-like event simulation chain re-uses many of the tools initially developed for the Baikal NT-200 project \cite{nt200}.
A simple neutrino generator code is used to simulate CC interactions of muon neutrino and muon anti-neutrino with nuclei in water.
The simulated neutrino energy spectrum and angular distribution follow the Bartol flux model \cite{Bartol1996}.
The neutrino-nuclei CC interaction cross sections are calculated using the CTEQ4M parton distribution functions \cite{CTEQ4M}.
The simulation covers the neutrino energy range from 10 GeV up to 100 TeV.
Energies above 100 TeV are excluded because they require an extrapolation of the Bartol model; their estimated contribution to the final event sample after all cuts (see Sect.~\ref{sect:analysis}) is $\approx 0.1$\%.
CC interactions of electron and tau neutrinos as well as all NC interactions are neglected for the track analysis studies.
The muon neutrino simulation proceeds as follows.
First, the incoming neutrino direction and energy are randomly drawn and the event kinematics (Bjorken y) is simulated.
Secondly, the neutrino interaction vertex is placed randomly within a parallelepiped, whose length is chosen so as to include the instrumented detector volume plus the typical maximum muon range for the given muon energy, with the two other dimensions big enough to include the instrumented volume plus additional 200 m around it (nearly 10 light absorption lengths).
The muon is propagated towards and through the instrumented volume using MUM v1.3u \cite{MUM}.
The hadronic shower created by the neutrino interaction is simulated using a simplistic parameterized shower model.
The propagation of the resulting Cherenkov light in water is modelled assuming a wavelength-dependent absorption model based on in-situ measurements.
Light scattering is ignored as a sub-dominant effect.
The detection of the Cherenkov photons by the optical module is simulated 
using a parameterized angular acceptance model,
taking into account the wavelength dependence of the PMT quantum efficiency and the transparency of the glass and the optical gel.
The OM angular acceptance model is based on a historical laboratory measurement \cite{ang_acc},
with an {\it ad~hoc} correction for the angular extent of the acceptance curve applied as follows: $f(x) \to f(0.9 \cdot x))$, where $x = cos(\alpha)-1$, and $\alpha$ is the angle between the OM optical axis and the incident light direction.
This choice of the correction factor allows us to better reproduce the observed zenith angle distribution of atmospheric muons,
while also bringing the acceptance model in close agreement with more recent laboratory measurements.
The hardware trigger is simulated taking into account the values of the trigger thresholds measured individually for each detector channel in-situ.
Likewise, random noise hits are added to the simulated events taking into account the actually measured average noise rate in the relevant period for each detector channel.
The channels which are missing in real data (see Table~\ref{table:dataset}) are masked out in the simulation.
In order to cover any possible mis-calibrations of the hit timing, which may be still present in real data due to the preliminary status of the used calibrations, a mis-calibration error of 5~ns has been artificially added to the MC.
Additionally, a 30\% charge measurement error has been artificially added.
For the purposes of this study each detector cluster is simulated separately.

The atmospheric muon bundle simulation uses CORSIKA~5.7 \cite{CORSIKA} to model the Cosmic Ray (CR) interactions in the atmosphere and to propagate the secondary particles down to the lake surface level.
The simulation incorporates the QGSJET model of hadronic interactions \cite{QGSJET}.
For the CR composition, we use a multi-component model based on KASCADE data \cite{Wiebel_Sooth_1999}.
The simulation covers the energy range of the primary CR nuclei from 240 GeV up to 20~PeV per nucleon; this energy range contains $>99$\% of all atmospheric muon events recorded with Baikal-GVD.
The propagation of the muons from the lake surface down to and through the detector volume is handled with MUM.
In order to optimize the simulation processing speed, the same CORSIKA events are re-used multiple times, varying their initial positions relative to the detector volume, as well as rotating the muon bundles around their axes.
Thanks to this scheme, a $\approx 1$~yr equivalent livetime of MC atmospheric muons has been accumulated and used for this analysis.
Further simulation steps for the atmospheric muon events are the same as for neutrinos.

\section{The track reconstruction algorithm}
\label{sect:reco}

The muon track reconstruction algorithm includes a hit selection procedure and a $\chi^2$-like track fitter.
The algorithm takes as input a list of PMT hits recorded in a 5~$\mu$s time window.
The hit information includes 
the hit time and charge, both obtained from an analysis of the recorded PMT signal waveform.
Each event is assumed to have only one muon track.
The hit with the highest charge deposition is used as a seed for further hit selection.
The following causality conditions are enforced for the other hits with respect to the seed hit:

\begin{equation}
\mid\Delta t\mid \, < \frac{r}{v} + \Delta t_{max},
\label{eq:1}
\end{equation}

\begin{equation}
\mid\mid\Delta t\mid \, - \, \frac{r}{c}\mid \, < \frac{r_{max}}{c},
\label{eq:2}
\end{equation}

where $\Delta t$ is the difference in the hit detection times between the seed hit and the hit in question, $r$ is the distance between the two PMTs where the two hits were detected, $c$ is the speed of light in vacuum, $v$ is the group velocity of light in Baikal water,
 and $\Delta t_{max}$ and $r_{max}$ are tunable parameters controlling the strictness of the causality conditions.
Condition \ref{eq:1} uses the fact that the Cherenkov light front propagates with a known speed and, in absence of strong light scattering, the Cherenkov photons cannot be delayed significantly behind the direct Cherenkov light front.
Condition \ref{eq:2} is effective when the distance between the two hits is substantially larger than the light absorption length: in that case the time difference between the hits becomes dominated by the muon propagation time.
In the following analysis we use $\Delta t_{max}$ = 10~ns and $r_{max}$ = 30~m.
With these settings, the hit selection provides a $\approx 75$\% purity for atmospheric neutrino events under typical detector conditions.

Using the selected hits, a preliminary muon track parameter estimation is performed.
First, a reference point is defined as the position of the seed hit shifted downwards along the vertical axis by 1 m.
The time of the seed hit is taken as the time of the muon crossing the reference point. 
Next, the direction of the track is estimated using a time-ordered list of hits, as follows:

\begin{equation}
\vec{R} = \sum_{j>i} w_{ij}(\vec{R_{j}}-\vec{R_{i}}),
\end{equation}

where $\vec{R_{i}}$ and $\vec{R_{j}}$ are the coordinate vectors of the $i$-th and $j$-th hit ($t_{j}>t_{i}$), respectively, and $w_{ij}$ is the sum of charges of hits $i$ and $j$, provided that the two hits are located on two different detector strings (otherwise $w_{ij}$ is set to zero).
The track direction is then represented by a dimensionless unit vector $\vec{k}=\vec{R}/\vert \vec{R} \vert$.
The described simple procedure provides a median angular resolution of $\approx 5^{\circ}$ (under low noise conditions).
The estimated track position and direction are used to refine the hit selection.
For this, the compatibility of the hit with the track is tested using
the residual time with respect to direct Cherenkov light propagation from the muon ($t_{res}$), the distance from the track ($\rho$), and a probability estimate for the hit to be observed at the OM ($p_{hit}$).
If the obtained values fail selection criteria, the hit is excluded.
The procedure is repeated 8 times with the selection criteria evolving from very loose to tight, with the tightest cuts being $t_{res}<100$~ns, $\rho<100$~m, and $p_{hit}>10^{-4}$.
This iterative procedure mitigates the impact of noise hits on the track direction estimation.
As a result, the contribution of noise hits to neutrino events is reduced to $\sim$1\%.

Using the refined hit selection, the muon position (at a given time $t_0$)
and flight direction (expressed with spherical coordinates $\theta$ and $\phi$) are reconstructed by means of minimisation of the quality function 

\begin{equation}
Q (x, y, z, \theta, \phi, t_0) = \chi_t^2 + Q_r.
\end{equation}

Here $\chi_t^{2}$ is the chi-square sum of the time residuals with respect to direct Cherenkov light from the muon:
\begin{equation}
\chi_t^{2} = \sum_i {(t_i^{exp} - t_i^{th})^2 \over \sigma^2},
\end{equation}
where $t_i^{exp}$ is time of the $i$-th detected hit, $t_i^{th}$ is the corresponding expected time for direct Cherenkov light from the muon, and $\sigma$ is an estimate of the time measurement precision. In the following analysis we use $\sigma$ = 3 ns. 

$Q_r$ is defined as follows:

\begin{equation}
Q_r = w \sum_i {a_{0} q_{i}  \over \sqrt{a_{0}^2 + q_i^2}} \sqrt{d_1^2 + r_i^2},
\end{equation}

where $w$ controls the relative weight of $Q_r$ in $Q$ (we use $w = 0.2$ m$^{-1}$[p.e.]$^{-1}$), $q_i$ is the charge of the $i$-th hit, $r_i$ is the distance of the $i$-th hit from the track, and $a_0$ and $d_1$ are configurable parameters (we use $a_0$ = 50 p.e. and $d_1$ = 1 m). 
The composition of the $Q_r$ term is inspired by \cite{antares_fastReco}.
It is motivated by the fact that the light intensity in a Cherenkov cone falls with distance $r$ approximately as $1/r$ (ignoring light absorption). That is, $q_i r_i \propto const$. 
The presence of a relatively high charge hit at a large distance from the track will lead to a large contribution to $Q_r$, thus penalizing the respective track position hypothesis.
The purpose of $a_0/\sqrt{a_{0}^2 + q_i^2}$ is to limit the effect of hits with very large charge depositions, while $d_1$ regularizes the fitter behaviour near $r_i=0$.
The hit selection method, the structure and the parameters of the minimisation function were optimised for the best reconstruction and selection of upgoing muons in the low energy range (E $\lesssim$ 1 TeV).
Using a typical modern CPU, the reconstruction of one triggered event takes $\approx 10$~ms, favoring the use of this algorithm for real-time applications.

\section{Data Analysis and Results}
\label{sect:analysis}

\begin{figure}
  \centering
  \includegraphics[height=7.5cm]{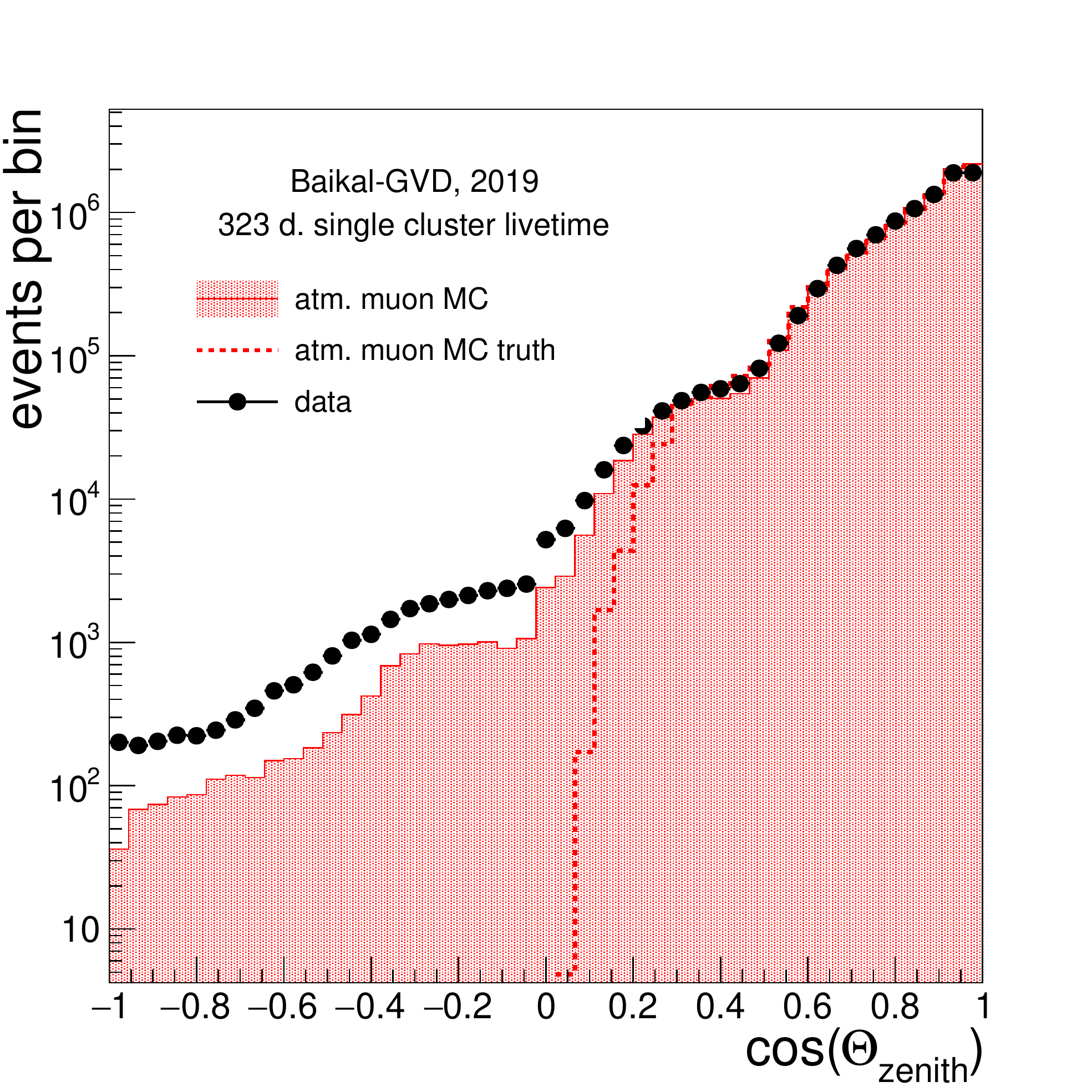}
  \caption{Zenith angle distribution of reconstructed tracks before quality cuts.
The Baikal-GVD 2019 data (five clusters with a combined livetime of 323 days) are shown by black points.
The MC prediction for atmospheric muons is shown by red filled histogram.
Additionally, the distribution of the true zenith angle of the MC events is shown (dashed red line).
 }
  \label{fig:track_analysis_results_1}
\end{figure}

The dataset described in Sect.~\ref{sect:dataset} has been 
processed with the track reconstruction algorithm described in Sect.~\ref{sect:reco}
after applying the necessary calibrations.
A minimal requirement of at least 8 hit OMs on at least two detector strings has been applied to ensure favorable conditions for accurate reconstruction of the zenith angle.
Additionally, a good fit convergence status is required.
This yields 9.8 million reconstructed events for the combined dataset from the 5 detector clusters.
The reconstructed zenith angle distribution of these events is shown in Fig.~\ref{fig:track_analysis_results_1}.
The experimental data is shown by black points.
The same reconstruction and analysis procedure has been applied to the simulated atmospheric muon sample (see Sect.~\ref{sect:mc}), resulting in the zenith angle distribution shown in Fig.~\ref{fig:track_analysis_results_1} in red.
According to the MC simulation, only a few percent of these events are single muons; the rest is caused by muon bundles.
Note that no fit quality requirements are applied at this point yet, and the majority of events reconstructed with cos $\theta<0.25$ are due to mis-reconstructed muon bundles.
Down to cos $\theta \approx 0.2$ we observe good agreement of experiment and MC.
Then, a discrepancy develops which reaches a factor of 3.5 for nearly vertically upward moving tracks (cos $\theta< 0.5$). This corresponds to an increased chance of mis-reconstruction for experimental data compared to MC.
It is currently unclear what is causing this discrepancy.
The effect is equally strong for each of the detector clusters.
In order to limit the impact of mis-reconstructed atmospheric muons,
in the following we focus on cos $\theta < -0.5$, 
where the background of mis-reconstructed atmospheric muons is small, 
and use very tight cuts to thoroughly eliminate the background.

\begin{figure}
  \centering
  \includegraphics[height=7.5cm]{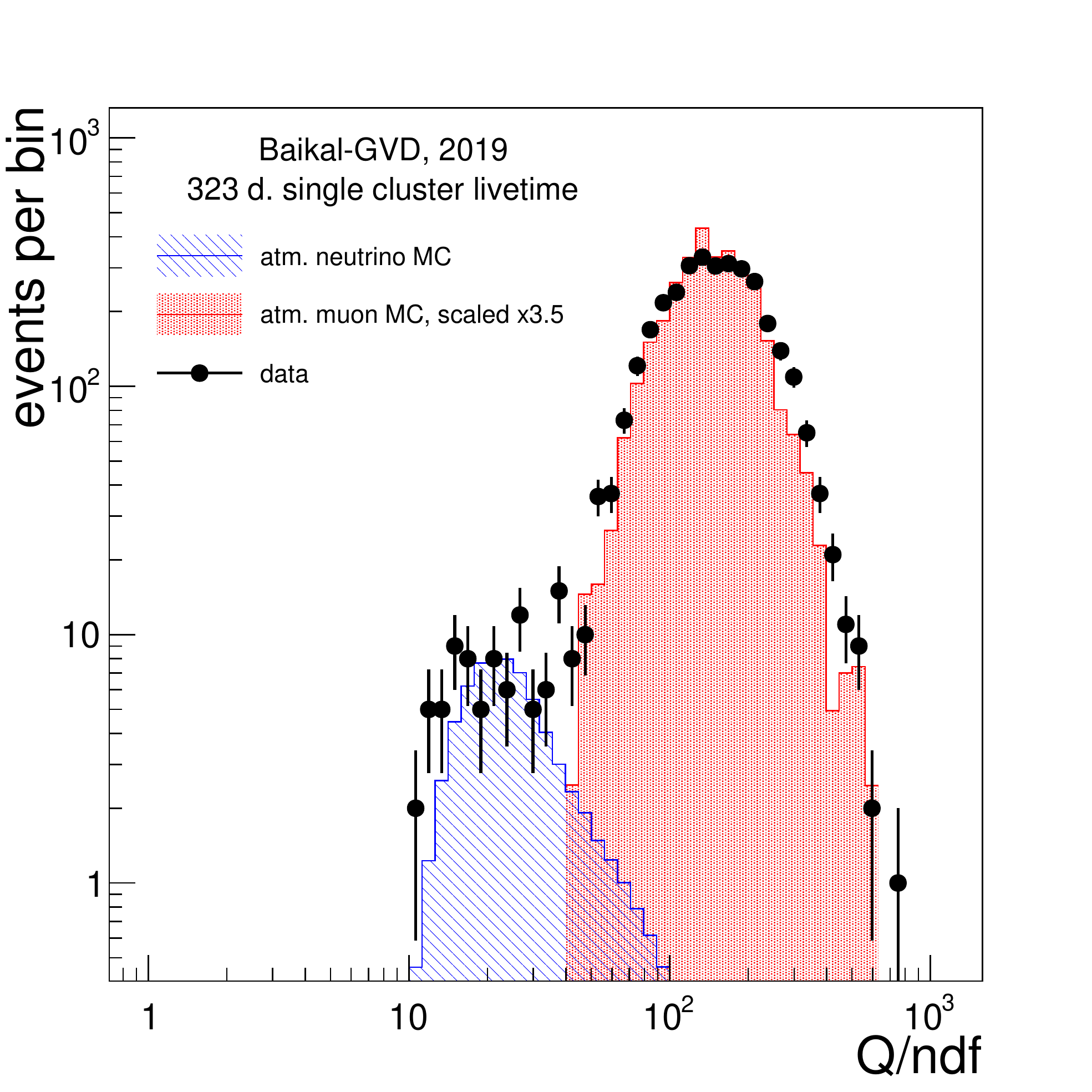}
  \caption{Distribution of the fit quality parameter for tracks reconstructed as upward-going with cos $\theta<-0.5$. 
The Baikal-GVD data are shown by black points with statistical error bars.
The MC predictions for atmospheric muons (scaled by a factor 3.5) and atmospheric neutrinos (not scaled) are shown in red and blue, respectively.
}
  \label{fig:track_analysis_results_2}
\end{figure}

The majority of the atmospheric muons can be suppressed by a cut on the zenith angle.
The remaining background of mis-reconstructed events can be further suppressed by tight cuts based on the fit quality and on other parameters.
Figure~\ref{fig:track_analysis_results_2} shows the distribution of the fit quality $Q$ divided by the number of degrees of freedom for events reconstructed with $\theta>120^\circ$ (cos $\theta<-0.5$). 
For this figure we have upscaled the atmospheric muon MC by the mentioned factor 3.5.
The data are shown in black.
The MC expectations for atmospheric muons and upgoing atmospheric neutrinos are shown in red and blue, respectively.
Obviously, the fit quality variable allows for efficient separation of the upgoing track-like neutrino events from the background of atmospheric muon events mis-reconstructed as upgoing. 
The shape of the fit quality distribution observed for experimental data is compatible with the MC prediction.

\begin{figure}
  \centering
  \includegraphics[width=8cm]{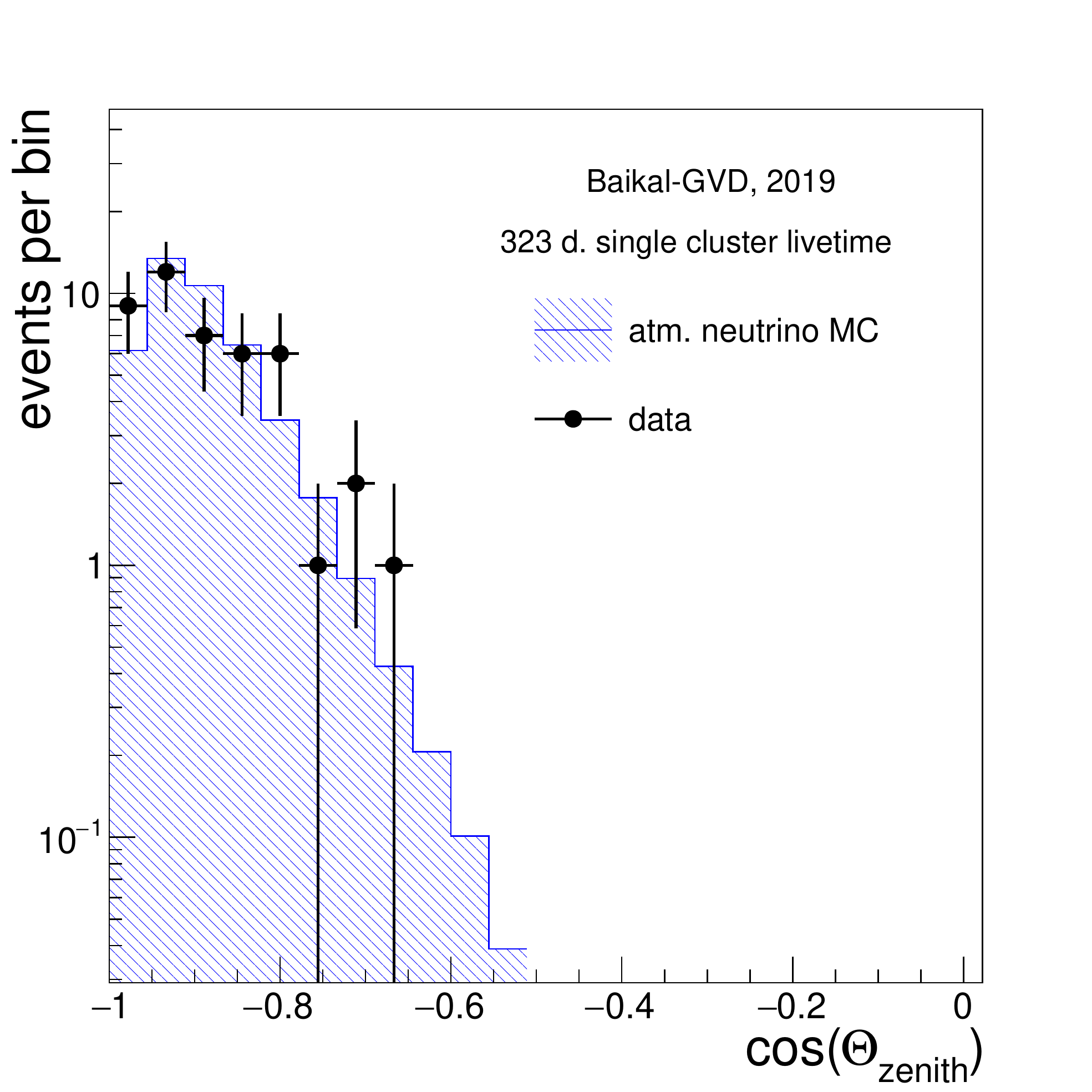}
  \includegraphics[width=8cm]{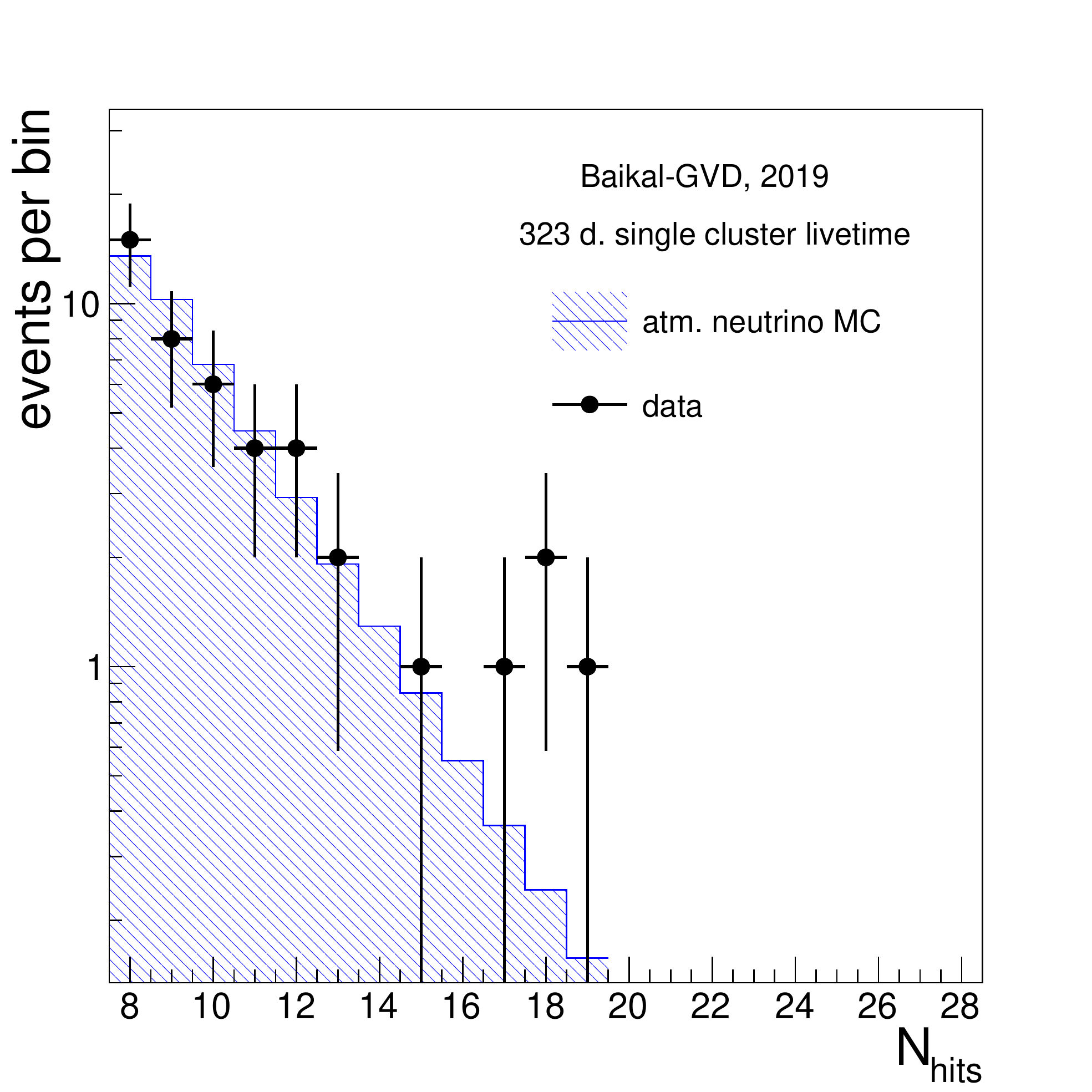}
  \caption{Results of the search for upgoing track-like neutrino events in the 2019 dataset (323 days single-cluster equivalent livetime).
           Left: zenith angle distribution.
           Right: distribution of the number of hits used in reconstruction.
           The black points and the blue filled area show the Baikal-GVD data and the MC prediction for atmospheric neutrinos, respectively.
           The expected number of background atmospheric muon events after the quality cuts is $\lesssim$1 and is therefore not shown.
 }
  \label{fig:track_analysis_results_3}
\end{figure}

After a cut on the fit quality parameter ($Q/ndf < 32$) we suppress the remaining background from atmospheric muons stepwise by cuts on further parameters. 
This includes the visible track length ($L > 42$~m), average distance from the track to the hit OMs ($\rho<18$~m), combined charge of the observed hits ($C > 18$~p.e.), estimated zenith angle error ($\theta_{err}<2^\circ$), as well as cuts on the width of the time residual distribution and additional hit likelihood variables.
The choice of cuts is highly conservative, aiming to ensure a high degree of background rejection under a wide range of optical noise conditions and in presence of sizeable systematic biases in Monte Carlo simulations.
Using these analysis cuts, a total of 44 neutrino candidate events were found in the experimental data while the expectation from the atmospheric neutrino MC simulation is $43.6$ $\pm \, 6.6$ (stat.) events, and the expectation for the atmospheric muon background is $\lesssim$1 event.
The median energy of the neutrino events, according to the MC simulations, is $\approx 500$ GeV.
The resulting zenith angle distribution of the neutrino candidate events is shown in  Fig.~\ref{fig:track_analysis_results_3}, left.
The distribution of the number of hits is shown in Fig.~\ref{fig:track_analysis_results_3}, right.
A good agreement with the MC prediction can be noted.
The apparent excess at $N_{hits}>16$ has a p-value of $\approx$ 0.05.

\section{Effective area and angular resolution}
\label{sect:performance}

\begin{figure}
  \centering
  \includegraphics[width=10cm]{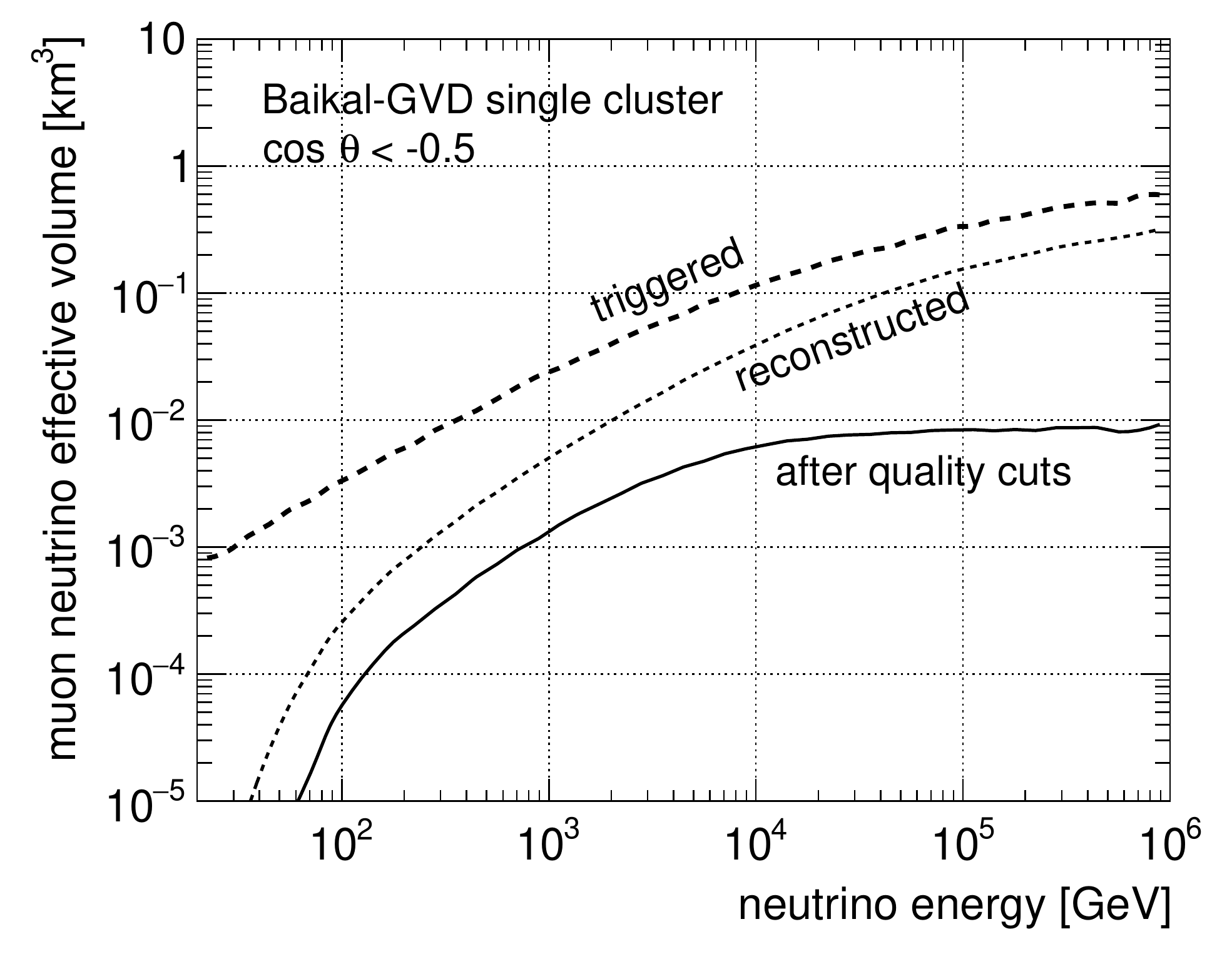}
  \caption{
Effective volume of a standard Baikal-GVD cluster for muon neutrino CC interactions at trigger level (thick dashed line), at reconstruction level using the $\chi^2$-based reconstruction (thin dashed line), and after the event selection cuts optimized for the atmospheric neutrino spectrum (solid line).
Shown are the average curves over the zenith angle range $\theta>120^\circ$ (upward-going).
 }
  \label{fig:eff_volume}
\end{figure}

The neutrino effective volume of one Baikal-GVD cluster using the reconstruction and analysis procedure described above is compared with the trigger-level effective volume in Fig.~\ref{fig:eff_volume}.
It can be noted that at high energies a $\approx$ 50\% reconstruction efficiency is reached,
however the efficiency declines towards lower energies, with a rapid drop at E $<$ 100 GeV.
The efficiency of the event selection cuts, on the contrary, declines towards high energies, making this event selection largely unsuitable for astrophysical neutrino searches.
At sub-TeV energies, where the majority of the atmospheric neutrino events are found, the analysis cut efficiency is found at a reasonable 25--30\% level.
A dedicated set of analysis cuts optimized for high energies is under development.
It is worth noting that at high energies the effective volume can exceed the detector instrumented volume by an order of magnitude.
This is due to the long muon range ($>10$ km) at high energies, which allows for the detection of neutrino events with initial vertices far outside the instrumented volume.

\begin{figure}
  \centering
  \includegraphics[width=8cm]{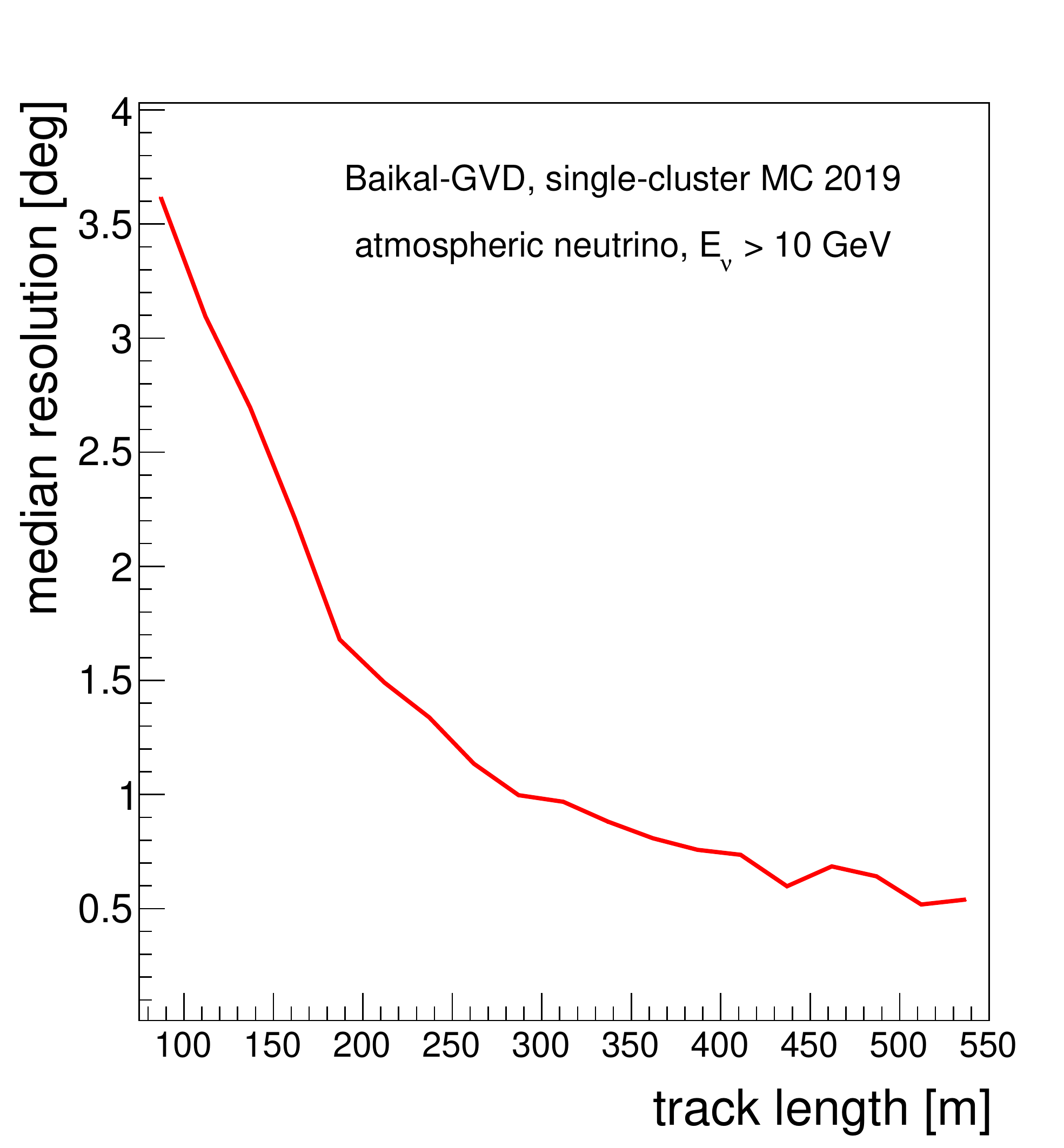}
  \caption{
Angular resolution of the $\chi^2$-based reconstruction method for muon neutrino observations using a single Baikal-GVD cluster after the event selection cuts, shown as a function of visible track length.
Shown are the median resolution values averaged over the atmospheric neutrino spectrum and zenith angle range $\theta>120^\circ$.
A 5 ns calibration accuracy is assumed.
 }
  \label{fig:ang_res}
\end{figure}

The projected angular resolution of the $\chi^2$-based reconstruction method for muon neutrino observations using a single Baikal-GVD cluster
is shown as a function of visible track length in Fig.~\ref{fig:ang_res}.
It can be noted that the resolution is inversely proportional to the track length.
A $0.5^\circ$ median resolution is reached for tracks with visible length $> 500$~m.
For the neutrino sample described in Sect.~\ref{sect:analysis} the expected median angular resolution is $1.6^\circ$.
The angular resolution is largely limited by the reconstruction method and the calibration uncertainties (which were assumed to be 5 ns in the simulation, see Sect.~\ref{sect:mc}).
Important improvements are therefore expected from improved calibrations, as well as from more advanced reconstruction techniques (likelihood-based track reconstruction, machine learning, etc.).
For comparison, the angular resolution in the Baikal-GVD cascade analysis is typically about $4^\circ$ \cite{Baikal_cascades}.

\section{Conclusion}
\label{sect:conclusion}
Baikal-GVD is a new TeV-PeV neutrino telescope under construction in Lake Baikal, Russia.
It consists of independently operated clusters of detector elements, with 8 strings per cluster and 36 optical modules per string.
The telescope has been collecting data in partial configurations since 2016.
As of April 2021, the detector consists of 8 clusters (64 strings, 2304 optical modules) with a total effective volume of 0.4 km$^3$.

We have developed a $\chi^2$-based reconstruction algorithm to reconstruct track-like events in Baikal-GVD.
The algorithm has been applied to a combined dataset of single-cluster events collected in 2019 from the first five operational clusters of Baikal-GVD.
Both the downgoing atmospheric muon flux and the upgoing atmospheric neutrino flux are studied.
The observations of atmospheric muons are found to be in fair agreement with expectations,
barring some unexplained difference in the rate of mis-reconstructed events.
The observed flux of atmospheric neutrino events, using a single-cluster analysis, is in good agreement with Monte Carlo predictions.
The algorithm is well suited for fast online data analysis applications
and is currently in use for physics analysis of the Baikal-GVD data.

\section*{Acknowledgements}
This work has been supported by the Ministry of Science and Higher Education of Russian Federation under the contract 075-15-2020-778 in the framework of the Large scientific projects program within the national project ``Science''.
We also acknowledge the technical support of JINR staff for the computing facilities (JINR cloud).

\end{document}